\documentclass[aps,pra,twocolumn,groupedaddress,showpacs]{revtex4}
\usepackage{epsfig}
\usepackage{graphicx}
\usepackage{amsfonts}
\usepackage{latexsym}
\usepackage{bm}

\def\8{\infty}

\def\d{\partial}

\def\undertext#1{\vtop{\hbox{#1}\kern 1pt \hrule}}

\def\VEV#1{\left\langle\,#1\,\right\rangle}
\def\tr{\hbox{tr}\,}

\def\pp#1{\frac{\partial}{\partial#1}}

\def\be{\begin{equation}}
\def\ee{\end{equation}}
\def\bea{\begin{eqnarray} & &}
\def\eea{\end{eqnarray}}

\def\rf#1{(\ref{#1})}

\def\rf#1{(\ref{#1})}

\def\cD{{\cal D}}
\def\sign{{\rm sign}}

\def\rfs#1{Eq.~\rf{#1}}

\def\nn{\nonumber}

\begin{document}


\title{Properties of the strongly paired fermionic condensates}


\author{J. Levinsen}
\author{V. Gurarie}

\affiliation{Department of Physics, CB390, University of Colorado,
Boulder CO 80309}


\date{\today}

\begin{abstract}
We study a gas of fermions undergoing a wide resonance $s$-wave
BCS-BEC crossover, in the BEC regime at zero temperature. We
calculate the chemical potential and the speed of sound of this
Bose-Einstein-condensed gas, as well as the condensate depletion, in the low
density approximation. We discuss how higher order terms in the low
density expansion can be constructed. We demonstrate that the
standard BCS-BEC gap equation is invalid in the BEC regime and is
inconsistent with the results obtained here. The low
density approximation we employ breaks down in the intermediate
BCS-BEC crossover region. Hence our theory is unable to predict how
the chemical potential and the speed of sound evolve once the
interactions are tuned towards the BCS regime. As a part of our
theory, we derive the well known result for the bosonic scattering
length diagrammatically and check that there are no bound states of
two bosons.
\end{abstract}


\maketitle


\section{Introduction}

In their seminal papers A. Leggett \cite{Leggett1980} and P.
Nozi\`eres and R. Schmitt-Rink \cite{Nozieres1985} studied a system
of spin-1/2 fermions with attractive short-ranged interactions in
the singlet channel at low temperature. If the interactions are
weak, the fermions form a Bardeen-Cooper-Schrieffer (BCS)
superconductor. If the interactions are made progressively stronger,
at some critical interaction strength a bound state of two fermions
becomes possible. These bound states are then bosons, which undergo
Bose-Einstein condensation (BEC). Studies by these and many other
authors demonstrated that as the interaction strength is increasing,
the BCS superfluid smoothly evolves into the  BEC superfluid,
without undergoing any phase transition in between.

Quite remarkably, Leggett, and Nozi\`eres and Schmitt-Rink afterwards,
found that the BCS gap equation, the mean-field equation which
describes the properties of the BCS superconductor, evolves under the
strengthening of the potential into the Schr\"odinger equation of a
bound pair of fermions. This happens despite the fact that the gap
equation is derived on the assumption of weak interactions between the
fermions and should in principle break down as the interaction
strength is increased. Interestingly, this observation was first made
as far back as in 1966 by V. N. Popov, \cite{Popov1966}, but his
contribution, at least within the context of the BCS-BEC crossover,
was not noticed until later. Similarly, the BCS-BEC crossover was
first studied in 1969 by D. M. Eagles \cite{Eagles1969} while the two
channel model \cite{Timmermans1999} was first studied by Yu. B. Rumer
\cite{Rumer1959}.

The fact that the gap equation describes the BCS superconductor
well, combined with the fact that the gap equation is also well
suited to describe the bound pairs of fermions on the opposite side
of the crossover, and taking into account that there is no phase
transition in between, allowed Nozi\`eres and Schmitt-Rink to
conclude that it may be used to interpolate between these two
regimes, when the potential is neither too weak nor too strong and
the fermions form an intermediate crossover BCS-BEC superfluid.

Interest in this subject was recently revived when the BCS-BEC
superfluid was obtained experimentally in the cold atomic systems
where the interactions can be tuned with the help of Feshbach
resonances
\cite{Timmermans1999,Holland2001,Timmermans2001,Hulet2003,Jin2004,Ketterle2004}.
Thus for the first time theoretical predictions regarding the BCS-BEC
crossover can be tested experimentally.

Given the lack of justification for the gap equation in the BEC
regime beyond predicting the binding energy of a pair of fermions,
the question remained whether it can be used to predict other
properties of the BEC condensate, such as the condensate depletion,
excitation spectrum and so on.


In this paper, we show that the gap equation is actually not valid
on the BEC side of the crossover. Although correctly predicting the
binding energy of the fermionic bound pairs, it fails to describe
the interactions of these dimers properly. As a result, if taken
literally, it incorrectly predicts the physical properties of the
BEC superfluid.

Indeed, it is generally believed in the literature
\cite{Leggett1980,Nozieres1985} that the transition between the BCS
and the BEC phases occurs without a phase transition making these
phases qualitatively the same phase.
Thus on a qualitative level,
it is possible to use the gap equation to obtain order of magnitude
estimates of the parameters of the BEC phase. However, if one is
interested in quantitative calculations, one needs to go beyond the
gap equation. Barring the exact solution of the problem, which most
likely is not possible here, the next best thing is to identify a
small parameter and do the calculations as an expansion in powers of
that parameter. The gap equation, when written on the BEC side of the
crossover, implicitly assumes that the small parameter is the
interaction strength between the bosons. This assumption is invalid;
it has been known for some time now that the Born approximation fails
to describe the interactions between the bosons properly
\cite{Petrov2005}.

In this paper, we identify as a small parameter the so-called gas
parameter of the Bose gas, and do all calculations as an
expansion in powers of that parameter.

The gap equation should be replaced by another equation on the BEC
side of the crossover. That equation, described below, cannot be
derived exactly, but can only be obtained as an expansion in powers of
the gas parameter. In the lowest order approximation, it allows us to
compute the chemical potential $\mu_b$, the speed of sound $u$ in the
superfluid, and the condensate depletion to be, at zero temperature,
\begin{eqnarray} \label{eq:results}
&&\mu_b = \frac{4 \pi n_b}{m_b} a_b,\,\,\, u^2=\frac{4 \pi n_b}{m_b^2}
a_b, \nn \\
&&n_{0,b} = n_b \left[1-\frac{8}{3}
\sqrt{\frac{n_b \, a_b^3}{\pi}} \right].
\end{eqnarray}
Here
\begin{equation} \label{eq:shlyap} a_b\approx 0.60 \,a
\end{equation}
is the scattering length of the bosonic dimers, whose approximate
relationship to the scattering length of a pair of fermions $a$ has
first been derived by Petrov, Salomon, and Shlyapnikov
\cite{Petrov2005}. The coefficient $0.60$ is approximate in the sense
of being determined only numerically, although the procedure to find
it is in principle exact. $n_b$ is the density of bosons, which in the
lowest order approximation in density can be replaced by $n/2$, where
$n$ is the density of the fermions. $n_{0,b}$ is the condensate
density. $m_b=2m$ is the mass of the bosons, where $m$ is the fermion
mass. $\mu_b$ is the chemical potential of the bosons, which can be
related to the chemical potential $\mu$ of the fermions by demanding
that $\mu_b-2\mu$ coincides with the binding energy released when the
boson is formed. Throughout this paper we adopt units in which $\hbar
= 1$.

Our results \rfs{eq:results} coincide with the behavior of the
dilute interacting Bose gas \cite{AbrikosovBook}. We would like to
emphasize that they are not exact, but are obtained as the lowest
order approximation in powers of the gas parameter $a
n^{\frac{1}{3}}$. It is possible to use the techniques described in
this paper and calculate the next order corrections to
\rfs{eq:results}. They will no longer necessarily coincide with the
next order corrections in the standard dilute Bose gas. We will not
calculate them here, however, and limit our discussion with merely
indicating how these higher order corrections can be obtained.

As the BEC condensate is tuned towards the crossover BCS-BEC regime,
$a$ increases reaching infinity at the unitary point which lies
somewhere in the intermediate BCS-BEC region. Thus the approximation
used in this paper breaks down as the intermediate regime is
approached. Our technique is unable to tell us anything about the
crossover regime, and we will not attempt to study it in this paper.

All calculations in this paper are done at zero temperature. We
think that our techniques can also be adopted to finite
temperatures, including studies of the critical superfluid transition
temperature, either in the large $N$ approximation or using some
other technique. See also Ref.~\cite{Andersen2004} for more
references on this subject. We will not attempt to do this in this
paper.

It is important to note that the interaction potential between
fermions is characterized not only by the scattering length $a$, but
also by the effective range $r_0$. Unlike $a$ which changes as the
interaction potential is tuned through the crossover, $r_0$ remains
roughly the same. If $\left| r_0 \right| \gg n^{-\frac{1}{3}}$, then
the fermions are said to be in the narrow resonance regime. In that
regime, the gap equation is a good approximation to the actual
physics of the BCS-BEC crossover everywhere \cite{Gurarie2005}.
However, the wide resonance crossover where $\left| r_0 \right| \ll
n^{-\frac{1}{3}}$, $\left|r_0 \right| \ll \left|a\right|$ is more
relevant for the current experiments, and it is the wide resonance
crossover which received most attention in the literature.
Everything we described above this paragraph pertains to the wide
resonance regime. In order to be certain that we work in the wide
resonance regime, we employ the one channel model which guarantees
that $r_0$ is vanishingly small.

An important part of the theory presented in this paper is the
diagrammatic derivation of \rfs{eq:shlyap}. It has come to our
attention that while our work was in progress, this was derived
diagrammatically, independently of us, by I.~V.~Brodsky, M.~Y.~Kagan,
A.~V.~Klaptsov, R.~Combescot, and X.~Leyronas, in
Ref.~\cite{Brodsky2005}. The techniques these authors employed to
arrive at \rfs{eq:shlyap} coincide with the ones discussed here.

Finally, we note that a problem similar to the one studied in this
paper was examined by L. V. Keldysh and A. N. Kozlov in the context
of excitons in Ref.~\cite{Keldysh1968}. Those authors, however,
concentrated on the case of Coulomb interactions of interest for the
physics of excitons, while we are interested in short ranged
interactions, which have quite  a different physics. In particular,
the universal results \rfs{eq:results} do not hold for their system.
Additionally, they, as well as Ref.~\cite{Popov1966}, used an
operator version of perturbation theory which is hard to generalize
beyond the lowest order. In contrast, our theory is based on a
functional integral and is easily generalizable to arbitrary order.
It is still remarkable that these authors understood the breakdown
of the gap equation in the BEC regime long before this became an
important issue in the BCS-BEC crossover literature.

The rest of the paper is organized as follows. In section
\ref{sec:one} we introduce the formalism and discuss how the
calculations of interest to us can be performed. In section
\ref{sec:two} we set up a perturbative expansion in powers of the gas
parameter. In the next section \ref{sec:three} we perform the actual
calculations and derive \rfs{eq:results}.  Section \ref{sec:BCS}
discusses an alternative derivation of the results of section
\ref{sec:three} and the origin of the breakdown of the BCS-BEC gap
equation. Finally, in section \ref{sec:scat}, which is followed by
conclusions, we compute the scattering amplitude between bosons and
the bosonic scattering length \rfs{eq:shlyap}, a necessary step in the
calculations performed in this paper. Some of the technical details of
the calculations are discussed in the appendices.

\section{Formulation of the problem}
\label{sec:one}

Consider a gas of spin-1/2 fermions, interacting with some short
ranged interaction in the $s$-wave channel. The most straightforward
way to study this gas is by means of a functional integral
\begin{equation} \label{eq:partst}
Z=\int \cD \bar \psi_\uparrow \cD \psi_\uparrow \cD \bar
\psi_\downarrow \cD \psi_\downarrow ~e^{iS_f},
\end{equation}
where $S_f=S_0+S_{\rm int}$ is the action consisting of the free
part
\begin{equation} \label{eq:S}
S_0=\sum_{\sigma=\uparrow,\downarrow} \int d^3x~ dt~ \bar
\psi_{\sigma} \left( i \pp{t}+\frac{1}{2m}\frac{\d^2}{\d {\bf
x}^2}+\mu \right) \psi_\sigma
\end{equation}
and the interaction part \begin{equation} \label{eq:Sint} S_{\rm
int} = \lambda \int d^3 x~dt~ \bar \psi_{\uparrow} \bar \psi_{
\downarrow} \psi_{\downarrow} \psi_{\uparrow}.
\end{equation}
This term describes interactions which happen at one point in space.
We need to remember that by itself such an interaction is unphysical
and has to be supplemented by a momentum cutoff $\Lambda$.
$\lambda>0$ to reflect that we choose an attractive interaction
potential.

As is standard in the treatment of superconductivity, we introduce
the Hubbard-Stratonovich field $\Delta$, which allows us to rewrite
\rfs{eq:partst} in the equivalent way
\begin{equation} \label{eq:part1}
Z=\int \cD \bar \psi_\uparrow \cD \psi_\uparrow \cD \bar
\psi_\downarrow \cD \psi_\downarrow \cD \Delta \cD \bar
\Delta~e^{iS},
\end{equation}
where the action $S$ is now given by $S=S_0+S_{\rm HS}$, and
\begin{equation} \label{eq:HS}
S_{\rm HS}= -\int d^3x~ dt~\left[ \frac{1}{\lambda} \bar \Delta
 \Delta +
\left( \Delta \bar \psi_{\uparrow} \bar \psi_{\downarrow} + \bar
\Delta \psi_{\downarrow} \psi_{\uparrow} \right)
\right].\end{equation}
The total action $S$ is now quadratic in
fermions, so the fermions can be integrated out. This results in the
following partition function
\begin{equation} \label{eq:part}
Z=\int \cD \Delta \cD \bar \Delta~e^{iS_\Delta},
\end{equation}
where the effective action $S_{\Delta}$ is given by
\begin{eqnarray}
\label{eq:Seff1} S_{\Delta}&=&-i~\tr \log \left( \matrix { \omega^+
-\frac{{\bf p}^2}{2m}+\mu & -\Delta \cr - \bar \Delta & \omega^+ +
\frac{{\bf p}^2}{2m}-\mu } \right) \cr & & -\frac{1}{\lambda}
 \int d t d^3 x~\bar \Delta \Delta,
\end{eqnarray}
where $\omega^+=\omega+i0\, \sign~\omega$. The action $S_\Delta$ and
the functional integral \rfs{eq:part} represent the starting point
in our theory, although sometimes it is also advantageous to keep
the fermions explicitly, as in \rfs{eq:part1} and \rfs{eq:HS}.

In the modern literature on Feshbach resonances, the system described
in Eqs.\rf{eq:partst}, \rf{eq:S}, and \rf{eq:Sint} is often referred
to as the one channel model. The interactions between fermions in this
model are such that their effective range $r_0$ is vanishingly small
(in the limit of a large momentum cutoff $\Lambda$). And indeed, the
results described in this paper are applicable to the limit $|r_0| \ll
n^{-{\frac{1}{3}}}$, $\left| r_0 \right| \ll \left|a \right|$ only,
thus the one channel model is a natural starting point for our
study. Alternatively, one could also study the two channel model
\cite{Timmermans1999}. The two channel model involves an additional
parameter which allows control of the value of $r_0$ independently of
$a$. However, if we chose to work with the two channel model, to
ensure the smallness of $r_0$ we would need to study the limit of
infinitely strong coupling within that model. In that limit, the
distinction between the two channel and one channel models
disappear. This is explained in more detail in Appendix
\ref{AppendixTC}. Because of this equivalence, we will not study the
two channel model any further in this paper.

The standard approach at this stage is to find the extremum of the
action $S_\Delta$ with respect to the field $\Delta$. Assuming that
the extremum of the action is when the field $\Delta$ takes some
constant value $\Delta_0$, we find the BCS-BEC gap equation
\begin{equation}
\frac{1}{\lambda} =\frac{1}{2} \int \frac{d^3 p}{(2 \pi)^3}
\frac{1}{\sqrt{ \left(\frac{ p^2}{2 m} -  \mu \right)^2 + \Delta_0^2
}}.
\end{equation}
The integral in the gap equation is up to momenta of the order of
the cutoff $\Lambda$. It is advantageous to trade the interaction
strength $\lambda$ for the fermionic scattering length $a$ using the
relation
\begin{equation}\label{eq:aforlambda}
a=\left(-\frac{4 \pi}{m \lambda} + \frac{2
\Lambda}{\pi}\right)^{-1}.
\end{equation}
This relation is fairly standard. For completeness, we include its
derivation in Appendix \ref{AppendixB}. Then the BCS-BEC gap
equation acquires the form
\begin{equation} \label{eq:BCSBEC}
-\frac{m}{4 \pi a} =\frac{1}{2} \int \frac{d^3 p}{(2 \pi)^3} \left [
\frac{1}{\sqrt{ \left(\frac{ p^2}{2 m} - \mu \right)^2 + \Delta_0^2 }}
- \frac{2 m}{p^2} \right].
\end{equation} The integral over momentum $p$ is now convergent and
can be extended to infinity.

If $a$ is negative and $|a| n^{\frac{1}{3}} \ll 1$ (this corresponds
to the weak attraction), then we say that the gas of fermions is in
the BCS state. In this regime, the gap equation has been
investigated in detail in the literature devoted to superconductors.
It was shown that the fluctuations of $\Delta$ about its mean field
value $\Delta_0$, which solves \rfs{eq:BCSBEC}, are indeed small,
and the physics of the superconductors can be captured by solving
\rfs{eq:BCSBEC}.

As the interaction strength is increased, $a$ becomes more and more
negative, reaching negative infinity at the interaction strength
which corresponds to the threshold of bound state creation. For even
stronger potentials, $a$ becomes positive, first large and then it
gets smaller. In this regime there is no justification for using
\rfs{eq:BCSBEC}.

Yet Ref.~\cite{Nozieres1985} argued that  in the ``deep BEC" regime
where $a n^{\frac{1}{3}} \ll 1$, \rfs{eq:BCSBEC} can be
reinterpreted as the Schr\"odinger equation of a pair of fermions,
with chemical potential playing the role of half of the energy. So
the gap equation then correctly predicts the formation of bosonic
dimers. Indeed, solving the gap equation in this regime by
anticipating that $\mu \ll -\Delta_0$,  we neglect $\Delta_0$  and
find within this approximation
\begin{equation} \label{eq:bind} \mu = -\frac{1}{2 m
a^2}.
\end{equation}
$1/ma^2$ is the binding energy of two fermions, as expected.

Nevertheless there is no reason to expect that the gap equation
holds beyond \rfs{eq:bind}. For example, one could expect that
attempting to use it to compute the deviation of $\mu$ from
\rfs{eq:bind} due to nonzero $\Delta_0$ in \rfs{eq:BCSBEC} would
lead to an incorrect relation between the boson chemical potential
$\mu_b=2\mu+1/(ma^2)$ and $\Delta_0$. On the other hand knowing the
proper relation between $\mu_b$ and $\Delta_0$ would allow us to
compute the physical properties of the BEC condensate, such as the
speed of sound. So there is a need to develop a reliable technique
which allows computation of these quantities.

In this paper we  investigate the properties of the system in the
low density ``deep BEC" regime, that is, in the regime where $a$ is
positive and small, so that $a n^{\frac{1}{3}} \ll 1$.  To this end,
we will calculate the normal and anomalous propagators of the
bosonic $\Delta$ field using a diagrammatic expansion. In doing so,
we will keep only the diagrams lowest in the density of the system.
However, we will not assume that the interactions are weak, or in
other words, that the interactions can be treated in the Born
approximation.

The fact that the Born approximation is not applicable to the
bosonic dimers has been known for quite a while. Indeed, the dimer's
scattering length in the Born approximation is $a_b=2 a$
\cite{Melo1993}, while the correct scattering length is given by
\rfs{eq:shlyap}. Although first derived in Ref.~\cite{Petrov2005} by
solving the four-body Schr\"odinger equation, it can also be derived
by  summing all the diagrams contributing to the scattering exactly.
In this paper we
 perform this summation and derive \rfs{eq:shlyap}
diagrammatically in section \ref{sec:scat}. This allows us to
recognize $a_b$ when the diagrams contributing to it appear within
the diagrams for the normal and anomalous bosonic propagators in our
theory.

Once the bosonic propagators are calculated, we will impose the
condition that they have a pole at zero frequency and momentum,
representing the sound mode of the condensate. This will give us a
relation between the chemical potential $\mu$, the scattering length
$a$, and the expectation value $\Delta_0$ of the field $\Delta$,
which will replace the gap equation \rfs{eq:BCSBEC}. In the dilute
Bose gas literature, this  is usually referred to as the
Hugenholtz-Pines relation \cite{Hugenholtz1959}. This equation
replaces the gap equation in our theory.

If instead of this proper procedure we had used the gap equation, we
would have arrived at \rfs{eq:results} with $a_b=2a$, which is the
Born approximation to the dimer scattering length. Thus the gap
equation fails because it cannot treat the interaction between
dimers beyond Born approximation.

In order to be able to determine the unknown quantities $\mu$ and
$\Delta_0$, one also needs another equation. Its role is usually
played by the particle number equation, which states that the total
particle density is equal to $n$,
\begin{equation} \label{eq:pne}
n=\frac{1}{V} \int d^3 x~\left[\VEV{\bar \psi_\uparrow
\psi_\uparrow}+\VEV{\bar \psi_\downarrow \psi_\downarrow} \right],
\end{equation}
where $V$ is the volume of the system. In its most naive
incarnation, we calculate the particle density by assuming that the
field $\Delta$ does not fluctuate and is equal to its mean value
$\Delta_0$. Then we find
\begin{equation} \label{eq:pne1}
n=\int \frac{d^3p}{(2\pi)^3}
\left[1-\frac{\frac{p^2}{2m}-\mu}{\sqrt{\left(\frac{p^2}{2m}-\mu\right)^2+\Delta_0^2}}
\right].
\end{equation}
We are going to show that this equation is indeed correct in the
lowest order approximation in density. However, if one wants to go
beyond the lowest order approximation, \rfs{eq:pne1} acquires
nontrivial corrections.

\section{Diagrammatic Expansion}
\label{sec:two}

To set up a diagrammatic expansion, we first write down free
propagators which follow from \rfs{eq:HS} or \rfs{eq:Seff1}, taking
into account that we work in the BEC regime where $\mu<0$. The
propagator of the fermions is given by
\begin{equation}
G_0(p,\omega)=\frac{1}{\omega-\frac{p^2}{2m}+\mu+i0}.
\end{equation}
Note that since  $\mu<0$, the propagator is retarded.

The propagator of bosons is, naively, equal to $-\lambda$. However,
we also need to include in it the self energy correction due to
fermionic loops. In other words, we need to expand the effective
action $S_\Delta$ up to quadratic order in $\Delta$ and the
corresponding term in the expansion corrects the propagator. This
expansion is performed in Appendix \ref{AppendixB}. We find, again
taking into account that $\mu<0$,
\begin{equation} \label{eq:D0}
D_0(p,\omega)=\frac{4 \pi}{m} \frac{1}{a^{-1}-\sqrt{m} \sqrt{-
\omega+\frac{p^2}{4m}-2 \mu-i0}}.
\label{eq:bosonprop}
\end{equation}
The bosonic propagator is also retarded, as any propagator of free
bosons must be at zero temperature \cite{AbrikosovBook}.

The fact that all the free propagators are retarded crucially
simplifies the diagrammatic expansion. Note that in the BCS regime
where $\mu>0$ this simplification would not have taken place.
Similar effect happens in the more standard dilute Bose gas
\cite{AbrikosovBook}.

The diagrammatic technique must also involve the lines beginning or
ending in the condensate. These are assigned the value $\Delta_0$.
Although $\Delta_0^2$ does not coincide with the density of
particles in the condensate, it is proportional to it. In turn, the
condensate density is always lower than the total density $n$. Thus
the expansion in powers of the density $n$ can be replaced by the
expansion in powers of $\Delta_0$. Since each line beginning or
ending in the condensate is assigned the value $\Delta_0$,  the
expansion in the density can also be understood as the expansion in
numbers of lines  beginning or ending in the condensate.

Now we would like to construct self energy corrections to the
propagators $G_0$ and $D_0$, to find the low density approximation
to the exact fermionic normal and anomalous propagators $G_n$ and
$G_a$ and exact bosonic normal and anomalous propagators $D_n$ and
$D_a$. We need to know $G_{n,a}$ to calculate the particle density
\rfs{eq:pne}. We need to know $D_{n,a}$ to relate the chemical
potential $\mu$ to other parameters in the theory.

We denote $\Sigma_{n,a}$ the normal and anomalous self energy
corrections to $G_{n,a}$. Keeping with tradition
\cite{AbrikosovBook}, we reserve the notation $\Sigma_{11}$ and
$\Sigma_{20}$ for the normal and anomalous self energy terms in
bosonic propagators $D_{n,a}$. We write
$\Sigma=\Sigma^{(1)}+\Sigma^{(2)}+\dots$, where $\Sigma^{(j)}$
denotes the contribution to $\Sigma$ from the diagrams with $j$ lines
beginning or ending in the condensate.

\begin{figure}[bt]
\includegraphics[height=1 in]{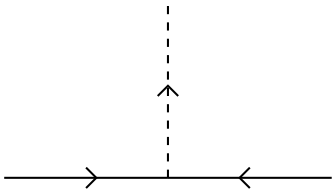}
\caption{The lowest order contribution in the low density
  approximation to the anomalous fermion self energy, $\Sigma_a$. The
  dashed line is a condensate line.}
\label{sigmaa}
\end{figure}
The only diagram with one condensate line which contributes to
$\Sigma_a$ is shown on Fig \ref{sigmaa}. We find
\begin{equation}
\Sigma_a^{(1)}=\Delta_0.
\end{equation}
The reason for the absence of any other diagrams at this order is
the absence of vertex corrections in our theory at $\mu<0$ when all
the propagators are retarded. Thus it is a feature of the BEC regime
only.

There are no normal self energy terms with one external line, thus
$\Sigma_n^{(1)}=0$. Computing the normal Green's function $G_n$ with
this one self energy correction is equivalent to computing the
correlation functions in \rfs{eq:pne} by assuming that $\Delta$ does
not fluctuate and is equal to $\Delta_0$. This gives \rfs{eq:pne1}.

\begin{figure}[bt]
\includegraphics[height=.7 in]{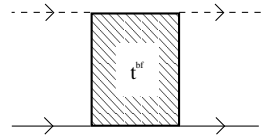}
\caption{The next to leading order contribution in the low density
  approximation to the normal fermion self energy.}
\label{sigmaan}
\end{figure}
No diagrams with two condensate lines contribute to the
anomalous self energy, and thus $\Sigma_a^{(2)}=0$. The
diagrams contributing to the normal self energy at this order
are shown in Fig. \ref{sigmaan}. The square block shown on this figure
is equal to the sum of all diagrams which contribute to boson-fermion
scattering. There is an infinite number of such diagrams, and they can
all be summed by solving the Lippmann-Schwinger integral
equation. This will be discussed in more details in section
\ref{sec:scat}. For now we denote the result of summation as $t^{\rm
bf}$. Then the contribution to the normal self energy
at this order is given by
\begin{equation}
\Sigma_{n}^{(2)}=\Delta_0^2 t^{\rm bf}.
\end{equation}


If needed, diagrams with three or higher number of external lines
can also be constructed.

\begin{figure}[bt]
\includegraphics[height=1.5 in]{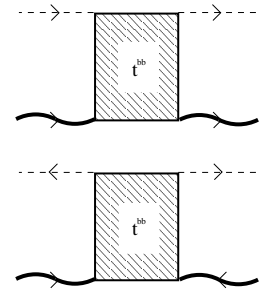}
\caption{The normal (top) and anomalous (bottom) self energy
  contributions to the bosonic propagator. The wavy lines are bosonic
  propagators.}
\label{sigma}
\end{figure}
Turning to the bosonic propagator, we find that there are no diagrams
contributing to $\Sigma_{11}$ and $\Sigma_{20}$ at the order of
$\Delta_0$. At the order of $\Delta_0^2$, the self energy diagrams are
shown on Fig \ref{sigma}.  The square box denotes the sum of all the
diagrams which contribute to the $T$-matrix of scattering of a boson
off another boson.  Just as those for $t^{\rm bf}$, these diagrams can
also be summed up using the appropriate Lippmann-Schwinger
equation. We denote the result of summation $t^{\rm bb}$. $t^{\rm bb}$
is calculated in section \ref{sec:scat}. Thus we find
\begin{equation} \label{eq:set}
\Sigma_{11}^{(2)} = 2 \Delta_0^2 t^{\rm bb}, \ \Sigma_{20}^{(2)}=
\Delta_0^2 t^{\rm bb}.
\end{equation}
The coefficient $2$ appears as the appropriate combinatorial factor.

\begin{figure}[bt]
\includegraphics[height=1.2 in]{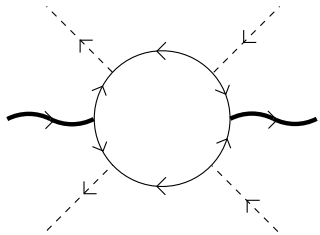}
\caption{A possible higher order contribution to the bosonic self energy.}
\label{highorder}
\end{figure}

It is also possible to consider diagrams higher order in $\Delta_0$
which contribute to the bosonic normal and anomalous self energy.
One such diagram, which contributes to $\Sigma_{11}^{(4)}$, is shown
 on Fig \ref{highorder}. There are infinitely many similar diagrams
 contributing to $\Sigma^{(4)}_{11}$ and $\Sigma_{20}^{(4)}$. We
 will not discuss these any further.

\section{Calculation of the speed of sound and the condensate depletion}
\label{sec:three}

Once the self energy terms are known, the procedure for calculating
the chemical potential, the speed of sound, and the condensate
depletion is fairly standard.

First of all, we calculate the normal and anomalous bosonic Green's
functions $D_{n,a}$, given $\Sigma_{11}$ and $\Sigma_{20}$. The
calculation involves solving the appropriate Dyson equation
\cite{AbrikosovBook} and the result is
\begin{eqnarray}
D_n(p,\omega)&=&
\frac{D_0^{-1}(-p,-\omega)-\Sigma_{11}(-p,-\omega)}{D(p,\omega)},
\cr D_a(p,\omega)&=&\frac{\Sigma_{20}(p,\omega)}{D(p,\omega)},
\end{eqnarray}
where
\begin{eqnarray}
D(p,\omega)&=&-\left[\Sigma_{20}(p,\omega) \right]^2+ \left(
D_0^{-1}(p,\omega) -\Sigma_{11}(p,\omega)\right)\times \cr && \times
\left( D_0^{-1}(-p,-\omega) - \Sigma_{11}(-p,-\omega) \right). \cr
&&
\end{eqnarray}
The Hugenholtz-Pines relation then takes the form
\begin{equation}
D(0,0)=0.
\end{equation}
In the lowest order in density, the self energy terms $\Sigma_{11}$
and $\Sigma_{20}$ vanish. Then the Hugenholtz-Pines relation reads
\begin{equation}
D_0^{-1}(0,0)=0.
\end{equation}
Recalling the definition of $D_0$, \rfs{eq:D0}, we immediately find
$\mu=-1/(2 m a^2)$, the same relation as the one which follows from
the gap equation, \rfs{eq:bind}.

In the first nonvanishing order  in density, the self energy terms
are given by \rfs{eq:set}. We are interested in the self energy at
zero momentum and frequency, thus $t^{\rm bb}$ also needs to be
computed at zero momentum and frequency. In section
\ref{sec:scat} we show how to calculate $t^{\rm bb}$ in vacuum by
solving the appropriate Lippmann-Schwinger equation. We, however,
need $t^{\rm bb}$ at a finite chemical potential $\mu$. Fortunately,
if
\begin{equation} \label{eq:impcond} \left| \mu+1/(2 m a^2) \right| \ll \left| \mu \right|,
\end{equation} then $t^{\rm bb}$ at a finite $\mu$ is approximately
given by its expression computed in vacuum at zero momentum and
frequency. In other words, it is proportional to $a_b$, where $a_b$
is the bosonic scattering length in vacuum, \rfs{eq:shlyap}. The
Lippmann-Schwinger equation can only be solved numerically, so we
only know the numerical value of $a_b$.

More precisely, $t^{\rm bb}$ at zero momentum and frequency, and at
chemical potential satisfying \rfs{eq:impcond} is related to the
scattering length $a_b$ as follows
\begin{equation} \label{eq:tbbfrac}
t^{\rm bb} = \frac{2 \pi a_b}{m Z^2},
\end{equation}
where $Z$ is the residue of the bosonic propagator $D_0$  at its
pole, or
\begin{equation} \label{eq:Z}
Z= \frac{8 \pi }{m^2 a}.
\end{equation}
We expect \rfs{eq:impcond} to hold as we expect $\mu$ to deviate
only slightly from its zero density value \rfs{eq:bind} in the low
density approximation employed here.

Armed with these relations, we introduce a bosonic chemical potential
$$
\mu_b=2\mu+\frac{1}{ m a^2} $$ and solve the Hugenholtz-Pines
relation in the next order in density
\begin{equation} \label{eq:ph}
\mu_b=\frac{ a \, a_b \, m \Delta_0^2}{4 }.
\end{equation}
There are actually two solutions of the Hugenholtz-Pines relation,
but one of them is well known to be unphysical \cite{AbrikosovBook}.

We remark that if we had instead decided to solve the gap equation
\rfs{eq:BCSBEC} by expanding its square root in powers of
$\Delta_0^2$, we would have obtained $\mu_b=\Delta_0^2 \,a^2 m/2$,
as if the bosonic scattering length were $a_b=2a$. This is
manifestly incorrect. Thus we see that the gap equation fails in the
BEC regime. In fact, it is easy to see exactly where the gap
equation breaks down, and what comes in its place. This analysis is
performed in section~\ref{sec:BCS}.

Now we use the particle number equation \rfs{eq:pne1} to relate
$\Delta_0$ to the particle density $n$. Within the lowest
approximation in density, we can substitute $\mu=-1/(2 m a^2)$ to
find
\begin{equation} \label{eq:pne2}
n=\frac{a m^2 \Delta_0^2}{4 \pi}.
\end{equation}
Using the next order approximation for the chemical potential would
only contribute to the further expansion of \rfs{eq:pne2} in powers
of $\Delta_0$, where, however, we would also need to take into
account the terms $\Sigma_{a,n}^{(2)}$. Fortunately this is not
needed for the calculations in the lowest order presented here.

Combining \rfs{eq:ph} and \rfs{eq:pne2} we find
\begin{equation}
\mu_b = \frac{\pi a_b  n}{m} =\frac{4 \pi a_b n_b}{m_b} ,
\end{equation}
where $m_b=2 m$ is the mass of the bosonic dimers and $n_b=n/2$ is
their density. This is the first of the results advertised in the
beginning of the paper in \rfs{eq:results}.

To find the speed of sound, we impose the condition
\begin{equation}
D(p,\omega)=0.
\end{equation}
This gives us a relation between $\omega$ and $p$. For small
$\omega$ and $p$, with the help of \rfs{eq:pne2} it reduces to
\begin{equation}
\omega^2 = p^2 \frac{4 \pi n_b a_b}{m_b^2}.
\end{equation}
Comparing with $\omega^2=u^2p^2$, where $u$ is the speed of sound,
gives the second result in \rfs{eq:results}.

Finally, we calculate the condensate depletion. The density of bosons
not in the condensate can be calculated as
\begin{equation} \label{eq:cd1}
\delta n_b = \frac{i}{Z} \lim_{t \rightarrow - 0} \int \frac{d
\omega \, d^3 p}{(2\pi)^4}~ D_n(p,\omega)\,  e^{-i \omega t}.
\end{equation}
The calculation of the integral in \rfs{eq:cd1} demands knowing the
full frequency and momentum dependence of the self energy terms
$\Sigma_{11}$ and $\Sigma_{20}$. This in principle can only be found
numerically, since the knowledge of full frequency and momentum
dependence of $t^{\rm bb}$ is required. However, in the limit of
small density where self energy terms are small, the relevant
frequency and momentum which contribute to the integral in
\rfs{eq:cd1} are in the range $\omega \sim \Delta_0^2$ and $p^2/(4m)
\sim \Delta_0^2$. Thus we expand $D_0^{-1}$ in powers of $\Delta_0^2
\sim |\omega+p^2/(2m_b)-\mu_b| \ll 1/(m a^2)$ to find the
approximate expression for the free bosonic propagator \rfs{eq:D0}
\begin{equation}
\frac{1}{Z} D_0(p,\omega) \approx \frac{1} {\omega-\frac{p^2}{4m} +
\mu_b+i0}.
\end{equation}
At the same time $Z \Sigma_{20}$ coincides with its expression for
the dilute interacting Bose gas. With this substitution,
$D_n(p,\omega)$ is equal to that of the dilute Bose gas, with the
scattering length $a_b$. Thus the standard formula for the
condensate depletion \cite{AbrikosovBook} holds, and it reads
\begin{equation} \label{eq:cd3}
n_{0,b}=n_b-\delta n_b \approx n_b \left[1-\frac{8}{3}
\sqrt{\frac{n_b \, a_b^3}{\pi}} \right],
\end{equation}
which is the final result in \rfs{eq:results}. We emphasize that this
is also an expansion in powers of $na^3$, and that next order terms
will likely depend on the functional dependence of the self energy
terms on the frequency and momentum.

Let us summarize what one needs to do to go beyond the lowest order
approximation discussed above.  The diagrams for the higher order
corrections to $\Sigma_{11}$ and $\Sigma_{20}$, such as the one shown
on Fig.~\ref{highorder}, have to be evaluated. There will be an
infinite number of such diagrams, and they can possibly be summed up
using a numerical technique similar to the one employed here to
calculate $t^{\rm bb}$. One also needs to calculate higher order
corrections to $\Sigma_a$ and $\Sigma_n$. In particular, at the order
of $\Delta_0^2$, $t^{\rm bf}$ has to be evaluated. Because of the
nonzero chemical potential present in the external lines of the
diagrams shown on Fig.~\ref{sigmaan}, $t^{\rm bf}$ will not reduce
simply to the scattering length $a_{bf}$, and so extra work will be
needed to find the value of
$\Sigma_n^{(2)}$. As pointed out by Hugenholtz and Pines in Ref.
\cite{Hugenholtz1959}, at each order one also needs to correct the
propagators such that a gapless sound mode is produced at this order.

Once the self energy terms are found in the desired order, one needs
to solve the Hugenholtz-Pines relation in the next order of
$\Delta_0^2$. This also involves taking into account the dependence
of $t^{\rm bb}$ on $\mu_b$, which was neglected here due to $\mu_b$
being zero up to terms of the order of $\Delta_0^2$. This allows to
find the chemical potential. Given the chemical potential, one can
also find corrections to the spectrum by looking at the poles of the
propagator. For that, the expansion of $t^{\rm bb}$ in powers of
momentum will also be needed. This expansion is also required in
order to find corrections to the condensate depletion, given at the
lowest order in \rfs{eq:cd3}.


\section{The origin of the failure of the BCS-BEC gap equation}
\label{sec:BCS}

An alternative method of deriving the results of the previous section
is presented here. A convenient way to derive the BCS-BEC gap equation
is by shifting $\Delta$ by $\Delta_0$ in \rfs{eq:part1} and
differentiating the resulting $Z$ with respect to $\bar\Delta_0$,
assuming that $\Delta_0$ gives an extremum of $\log Z$. This gives
\begin{equation} \label{eq:opb}
-\frac{\Delta_0}{\lambda} = \VEV{\psi_{\downarrow} \psi_{\uparrow}}.
\end{equation}
To derive the naive gap equation \rfs{eq:BCSBEC} one usually evaluates
the right hand side of \rfs{eq:opb} at constant
$\Delta=\Delta_0$. While this procedure works in the BCS regime, it
actually breaks down in the opposite BEC regime.

Indeed, on the BEC side of the crossover, it is more appropriate to
expand everything in powers of $\Delta_0$. So we instead compute the
right hand side of \rfs{eq:opb} in terms of diagrams with a fixed
number of condensate lines.

\begin{figure}[hbt]
\includegraphics[height=.5 in]{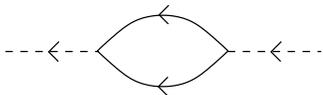}
\caption{The simplest diagram contributing to the gap equation
\rfs{eq:opb}.} \label{fig:cond1}
\end{figure}

The simplest diagram contributing to \rfs{eq:opb} is shown on
Fig.~\ref{fig:cond1}. There are two condensate lines shown on that
diagram. The line on the left is not an actual line. Rather it is
drawn simply for convenience, and the vertex where it is attached to
the diagram represents $\psi_{\downarrow} \psi_{\uparrow}$ whose
correlation function we are computing. The line on the right,
however, is the actual condensate line. Thus this diagram has one
condensate line and is proportional to $\Delta_0$.

If we use this diagram for the right hand side of \rfs{eq:opb}, we
derive \rfs{eq:BCSBEC} with the substitution $\Delta_0=0$. Thus this
reproduces the lowest order gap equation, and gives \rfs{eq:bind}
for $\mu$.

\begin{figure}[hbt]
\includegraphics[height=.9 in]{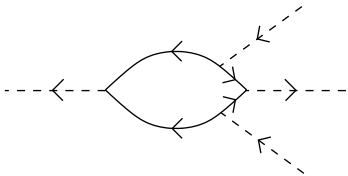}
\caption{The simplest diagram contributing to the gap equation at
  order $\Delta_0^3$.}
\label{fig:cond2}
\end{figure}

This is the only possible diagram with one condensate line. There are no
diagrams with two condensate lines either. Once we allow for three
external condensate lines, we find one possible diagram shown on
Fig.~\ref{fig:cond2}.

This diagram, when taken into account, gives the usual gap equation
\rfs{eq:BCSBEC} where the square root has been expanded up to linear
order with respect to $\Delta_0^2$. We know that solving
\rfs{eq:BCSBEC} at this order gives $a_b=2a$ \cite{Melo1993}. And
indeed, the diagram shown on Fig.~\ref{fig:cond2} is nothing but the
simplest process contributing to the boson-boson $T$-matrix, which is
the very first diagram on Fig.~\ref{fig:diagrams}. In other words, it
is the Born approximation for the boson-boson scattering.

\begin{figure}[hbt]
\includegraphics[height=.9 in]{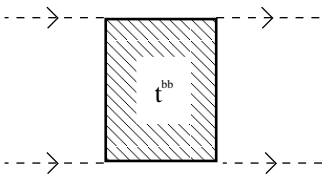}
\caption{The sum of all diagrams contributing to the gap equation at
  order $\Delta_0^3$.}
\label{fig:cond3}
\end{figure}

We know that the Born approximation breaks down in the BEC regime.
Instead, at this order we need to take all the diagrams with three
external lines, which are equal to $t^{\rm bb}$, as shown on
Fig.~\ref{fig:cond3}.
This gives the correct gap equation, with $2a$ being replaced by
$0.60\,a$.

To summarize, instead of using the naive gap equation \rfs{eq:BCSBEC}
which does not take into account fluctuations around $\Delta_0$, in
the BEC regime we use \rfs{eq:opb}, expanding the right hand side in
powers of $\Delta_0$. This procedure goes beyond the Born
approximation and reproduces the correct results in the BEC regime.


\section{Scattering of particles and bound states}
\label{sec:scat}
\begin{figure}[ht]
\includegraphics[height=0.36 in]{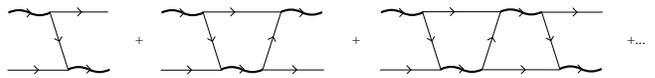}
\caption{Diagrams contributing to the $t$-matrix of fermion-boson
  scattering.}
\label{fig:firstfb}
\end{figure}
In this section we consider the scattering of a fermion by a bound
state of fermions and scattering between bound states of fermions.
All calculations proceed in vacuum so we set $\mu=0$ for this
entire section. As argued in section \ref{sec:three}, these vacuum
scattering processes are precisely those needed to calculate the
normal and anomalous fermion self energies at next to leading order
in the low density expansion and the lowest order boson normal and
anomalous self energies, respectively.

The scattering $t$-matrix consists of all diagrams with incoming and
outgoing lines corresponding to the scattering particles. Since we
work in vacuum, there is no hole propagation, or all the propagators
are retarded. These observations greatly simplify the possible
diagrams. In fact, had any of the propagators been advanced, there
would be no hope of summing the diagrams contributing to the
scattering processes exactly. Fortunately, when all the propagators
are retarded, we are able to sum all the diagrams using
Lippmann-Schwinger type integral equations. We emphasize that all the
calculations proceed in the regime of vanishing effective range $r_0$,
so the results of this sections are valid only for potentials where
$|a| \gg |r_0|$.

We first consider scattering of a fermion and a bound state of two
fermions, a bosonic dimer. This process was first solved in the zero
effective range approximation by Skorniakov and Ter-Martirosian
\cite{Skorniakov1956} in the related problem of neutron-deuteron
scattering. We include here a somewhat detailed description of this
scattering process in part because it is a useful exercise before
considering the more complicated case of scattering between two
bosons.
\begin{figure}[ht]
\includegraphics[height=1.85 in]{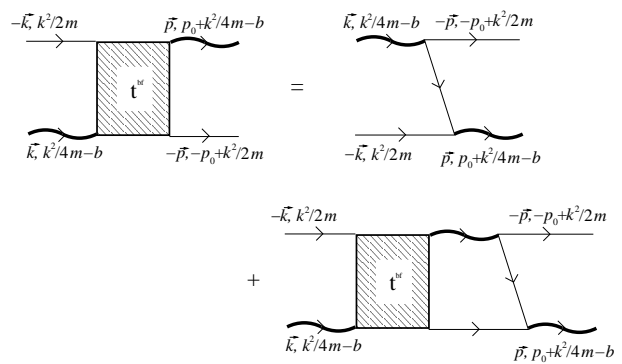}
\caption{The integral equation for the $t$-matrix of fermion-boson
  scattering.} \label{fig:fb}
\end{figure}

The first few diagrams contributing to the $t$-matrix are shown in
Fig. \ref{fig:firstfb}, where the external lines are not part of the
$t$-matrix. Even though all the propagators are retarded,  there is
still an infinite number of diagrams containing retarded propagators
only, and they all contribute at the same order. Indeed, from these
first examples it is clear that any such contribution to the
$t$-matrix will have $n\geq0$ boson lines of order $a/m$, see Eq.
(\ref{eq:bosonprop}), $2n+1$ fermion lines of order $1/E\sim ma^2$
and $n$ integrations of order $Ep^3\sim m^{-1}a^{-5}$. Thus each
diagram contributing to the $t$-matrix is of order $ma^2$.

There is no hope of computing these diagrams one by one and summing
them up. Instead, we derive an integral equation which their sum
satisfies.  Let the incoming boson and fermion have on-shell
4-momenta $(\vec k, k^2/4m-b)$ and $(-\vec k,k^2/2m)$, respectively,
and let the outgoing boson and fermion have 4-momenta $(\vec
p,p_0+k^2/4m-b)$ and $(-\vec p,-p_0+k^2/2m)$, respectively. Here $b$
is the binding energy of the dimer, $-1/(ma^2)\equiv -b$. The
$t$-matrix with these kinematics is denoted $t^{\rm bf}_{\vec
k}(\vec p,p_0)$. Then the $t$-matrix satisfies the integral equation
\begin{eqnarray}
t^{\rm bf}_{\vec k}(\vec p,p_0) & = & -G_0(\vec k+\vec p,-k^2/4m-b+p_0)
\nn \\ && \hspace{-2.2cm}
-i\int\frac{d^4q}{(2\pi)^4}t^{\rm bf}_{\vec k}(\vec
q,q_0) D_0(\vec q,k^2/4m-b+q_0)\nn \\ && \hspace{-2.2cm} \times
G_0(-\vec q,\frac{k^2}{2m}-q_0)G_0(\vec p+\vec
q,p_0+q_0-\frac{k^2}{4m}-b),
\label{eq:threebody}
\end{eqnarray}
which is depicted in Fig. \ref{fig:fb}. The minus sign in front of
the right hand side of the equation is due to fermionic
anticommutations. The reason for letting outgoing momenta be
off-shell is that this makes it possible to solve the integral
equation. The on-shell point has $|\vec k|=|\vec p|$ and $p_0=0$. It
is possible to integrate out the loop energy on the right hand side
in Eq. (\ref{eq:threebody}) by noting that $t^{\rm bf}_{\vec k}(\vec
q,q_0)$ is analytic in the upper half plane in $q_0$ which can be
easily seen by looking at the diagrams contributing to the
$t$-matrix. This integration sets $q_0=k^2/2m-q^2/2m$ and we thus
set $p_0=k^2/2m-p^2/2m$, to have the same dependence in the
$t$-matrix on both sides in the equation. Define $t^{\rm bf}_{\vec
k}(\vec p,p_0=k^2/2m-p^2/2m)\equiv t^{\rm bf}_{\vec k}(\vec p)$.

At low energies scattering is dominated by $s$-wave scattering, thus
Eq. (\ref{eq:threebody}) can be averaged first over directions of
$\vec k$ and then over directions of $\vec p$. The angular average of
$t^{\rm bf}_{\vec k}(\vec p)$ is denoted $t^{\rm bf}_{k}(p)$.  Then
the integral equation for the $t$-matrix of fermion-boson scattering
becomes
\begin{eqnarray}
t^{\rm bf}_k(p) & = & \frac m{2pk}\ln\frac{p^2+pk+k^2-mE}{p^2-pk+k^2-mE}
\nn \\ &&
+\frac1{\pi}\int^\infty_0dq\,\frac qp\ln\frac{q^2+qp+p^2-mE}
{q^2-qp+p^2-mE}
\nn \\ &&
\times \frac{t^{\rm bf}_k(q)}{a^{-1}-\sqrt m\sqrt{-E+3q^2/4m}}.
\label{eq:inteqbf}
\end{eqnarray}
Here, $E=3k^2/4m-b$ is the total energy which is assumed to be less
than zero. To calculate the scattering amplitude each external bosonic
leg has to be renormalized by the square root of the residue of the
pole of the bosonic propagator at the energy of the bound state,
$\sqrt Z=\sqrt{8\pi/(m^2a)}$, compare with \rfs{eq:Z}. The external
fermionic legs are free propagators and do not have to be
renormalized. The scattering amplitude is evaluated on shell and is
\begin{equation}
T^{\rm bf}(k) = Zt^{\rm bf}_{k}(k).
\end{equation}
The relationship between the scattering length and the scattering
amplitude is
\begin{equation}
T^{\rm bf}(0) = \frac{3\pi}ma_{bf}.
\end{equation}
Solving Eq. (\ref{eq:inteqbf}) it is found that $a_{bf}\approx1.18a$.

\begin{figure}[ht]
\includegraphics[height=1.85 in]{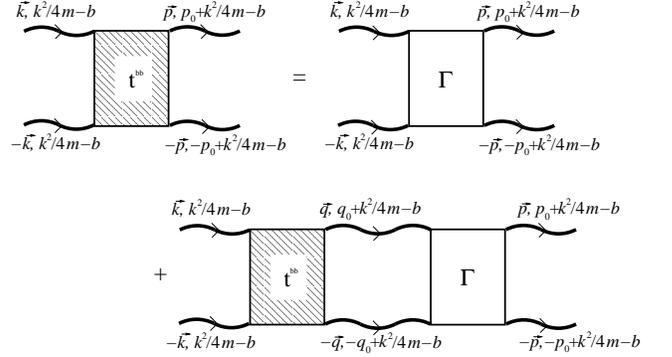}
\caption{The integral equation for the $t$-matrix of boson-boson
  scattering. $\Gamma$ is the sum of all two boson irreducible
  diagrams.} \label{fig:inteqgamma}
\end{figure}

We now turn to the scattering between two bosons where the bosons are
bound states of two distinguishable fermions. This process was first
solved by Petrov {\it et al} \cite{Petrov2005} by solving the quantum
mechanical 4-body problem. As we already mentioned, recently, while
the work reported here was in progress, the problem was solved in
Ref.~\cite{Brodsky2005} by using the same diagrammatic approach as we
employ below.

As in the case of fermion-boson scattering there are no condensate
lines internally in the diagrams and no hole propagation.

\begin{figure*}[hbt]
\includegraphics[height=0.6 in]{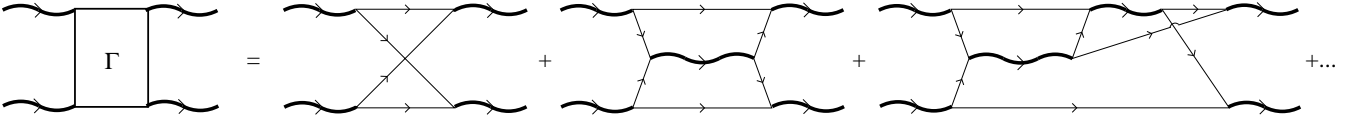}
\caption{The simplest diagrams contributing to the two boson
  irreducible diagram, $\Gamma$. The first diagram on the right hand
  side corresponds to $\Gamma^{(0)}$ defined in Eq. (\ref{gamma0}).}
  \label{fig:diagrams}
\end{figure*}

Consider the different diagrams which contribute to the $t$-matrix of
boson-boson scattering. The $t$-matrix consists of all possible
diagrams with two incoming and two outgoing bosons, the external legs
not included. The crucial point is that all of these diagrams are of
the same order, namely they are all proportional to $m^3a^3$ which we
will show below. This means that the diagrams contributing to the
$t$-matrix do not form a perturbation series and it is not adequate to
keep only the Born approximation to the $t$-matrix. Instead an
infinite number of diagrams must be taken into account. The summation
of all diagrams contributing to the $t$-matrix may be performed by
using an integral equation. In particular, the $t$-matrix may be built
from a series of two boson irreducible diagrams, which results in an
integral equation as shown in Fig. \ref{fig:inteqgamma}. Two boson
irreducible diagrams are understood to be diagrams which cannot be cut
in half by cutting two boson lines only.

Let the kinematics be as depicted on the figure with the incoming
energies and momenta chosen on-shell and the outgoing energies and
momenta off-shell to allow for the solution of the integral
equation. The on-shell condition is $|\vec p|=|\vec k|$ and
$p_0=0$. The $t$-matrix with kinematics as shown on the left hand side
of Fig. \ref{fig:inteqgamma} is denoted by $t^{\rm bb}_{\vec k}(\vec
p,p_0)$. The two boson irreducible diagram with $(\pm \vec q,E/2\pm
q_0)$ incoming 4-momenta and $(\pm \vec p,E/2\pm p_0)$ outgoing
4-momenta is given by $\tilde\Gamma_k(\vec q,q_0;\vec p,p_0)$. $E$ is
the total energy, $E=-2b+k^2/2$. The integral equation corresponding
to these kinematics is
\begin{eqnarray}
t^{\rm bb}_{\vec k}(\vec p,p_0) & = & \tilde\Gamma_k(\vec k,0;\vec p,p_0) \nn
\\ && \hspace{-1cm}+i\int\frac{d^4q}{(2\pi)^4}D_0(\vec q,E/2+q_0)D_0(-\vec q,
E/2-q_0)\nn \\ && \hspace{-1cm}\times\tilde\Gamma_k(\vec q,q_0,\vec p,p_0)t^{\rm
bb}_{\vec k}(\vec q,q_0).
\label{inteqgamma}
\end{eqnarray}
Thus far the integral equation does not depend on whether the
constituents of the dimers are bosons or fermions. Since the
scattering is dominated by $s$-wave scattering we average
Eq. (\ref{inteqgamma}) over directions of $\vec k$, then directions of
$\vec p$ and finally the integration over directions of $\vec q$ can
be trivially performed. Let the corresponding angular averages be
$t^{\rm bb}_k(p,p_0)$ and $\Gamma_k(q,q_0;p,p_0)$ and let $k\equiv |\vec k|$,
$p\equiv |\vec p|$, and $q\equiv |\vec q|$. Then the integral equation
becomes
\begin{widetext}
\begin{equation}
t^{\rm bb}_k(p,p_0) = \Gamma_k(k,0;p, p_0)+\frac{4i}\pi
\int
\frac{\Gamma_k(q,q_0;p,p_0)t^{\rm bb}_k(q,q_0) q^2dq\,dq_0}{1+
\sqrt{(q^2/4-E/2)^2-q_0^2}-\sqrt2\sqrt{q^2/4-E/2+
\sqrt{(q^2/4-E/2)^2-q_0^2}}},
\label{inteqgammas}
\end{equation}
\end{widetext}
which contains no remaining angular dependence and where the product
of bosonic propagators has been written explicitly using
Eq. (\ref{eq:bosonprop}). Since all diagrams are of the same order of
magnitude, it is convenient to let all momenta be measured in units of
$a^{-1}$ and energies in units of $b=1/(ma^2)$. $t^{\rm bb}$ and
$\Gamma$ are then measured in units of $m^3a^3$. For simplicity this
is the case in Eq. (\ref{inteqgammas}) and will be the case in the
remainder of this section.

\begin{figure*}[hbt]
\includegraphics[height=.69 in]{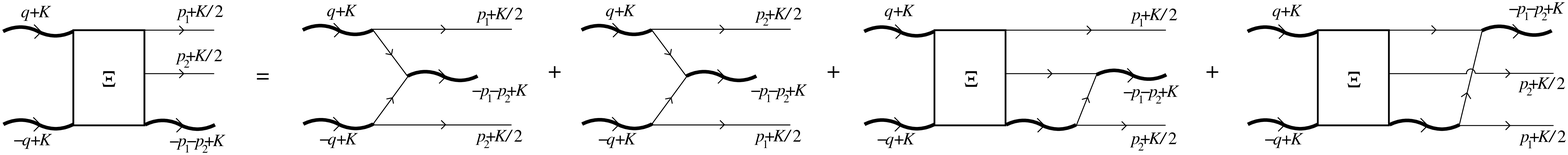}
\caption{Integral equation for the two boson irreducible diagram with
  two bosons coming in and one boson and two fermions going out.}
\label{fig:inteqxi}
\end{figure*}

The simplest two boson irreducible diagrams are shown in
Fig. \ref{fig:diagrams} where the external lines are not a part of
$\Gamma$. It is clear that there is an infinite number of such
diagrams. Letting $n$ count the number of bosonic propagators in the
diagram contributing to $\Gamma$, any such diagram has $4+2n$
fermionic propagators, each of order $ma^2$, $n$ bosonic propagators
of order $a/m$ and $n+1$ integrations of order $m^{-1}a^{-5}$. Thus
each of these diagrams is of order $m^3a^3$. To see that $t^{\rm bb}$
is also of order $m^3a^3$ note that any diagram contributing to
$t^{\rm bb}$ will contain $n+1$ factors of $\Gamma$, $2n$ bosonic
propagators, and $n$ integrations, resulting in any diagram
contributing to $t^{\rm bb}$ being proportional to $m^3a^3$.

Letting the first diagram on the left in Fig. \ref{fig:diagrams} equal
the $t$-matrix corresponds to the Born approximation which gives
$a_b=2a$. Using this same diagram for $\Gamma$ reproduces the result
of Ref. \cite{Pieri2000}, $a_b=0.78a$. However in both of these methods
an infinite number of diagrams of the same order as those considered
is ignored.

The sum of two boson irreducible diagrams may again be obtained
through solving an integral equation. The integral equation to be
solved is depicted in Fig. \ref{fig:inteqxi} where $\Xi$ denotes all
two boson irreducible diagrams with two incoming bosons, one outgoing
boson and two outgoing fermions. The two boson irreducible diagram
$\Gamma$ may then be obtained by tying together the two external
fermion lines.  Furthermore, since the two boson irreducible diagram
$\Gamma$ does not depend on angles, we choose to let $\Xi$ be averaged
over the direction of the incoming momentum.  Define a four vector
$K\equiv (\vec 0,E/2)$ which describes the incoming energy carried by
a boson. Then the integral equation corresponding to
the kinematics shown on the figure is
\begin{widetext}
\begin{eqnarray}
\Xi_k(q,p_1,p_2) & = & -\int\frac{d\Omega_{\vec q}}{4\pi}
\left[G_0 \left(\frac{K}{2}+q-p_1 \right)G_0\left( \frac K2 -q-p_2 \right)+G_0\left(\frac K2+q-p_2 \right)
G_0\left(\frac K2-q-p_1 \right)\right] \nn \\
&& \hspace{-3.1cm}-i\int\frac{d^4Q}{(2\pi)^4}G_0\left(\frac{K}2+Q
\right)G_0\left(\frac{K}2-Q-p_1-p_2\right)
\left[\Xi_k(q,p_1,Q)D_0(K-Q-p_1)+\Xi_k(q,Q,p_2)D_0(K-Q-p_2)\right].
\label{inteqxi}
\end{eqnarray}
\end{widetext}
The minus signs in front of both the first order terms and the higher
order terms are a result of anti-commuting fermions. The integral over
$d\Omega_{\vec q}$ is the averaging over angles of incoming momentum.
$\Gamma$ is then found from $\Xi$ by connecting the outgoing fermion
lines and integrating over the loop momentum. The precise relation
between $\Gamma$ and $\Xi$ is
\begin{eqnarray}
\Gamma_k(q;p) & = & \frac i2\int\frac{d^4q'}
{(2\pi)^4}G_0(K/2+q')G_0(K/2+p-q') \nn \\
&& \times \Xi_k(q,q',p-q').
\label{gammaxi}
\end{eqnarray}
It is not essential to include both lowest order contributions to
$\Xi$ in Eq. (\ref{inteqxi}), the only difference if only one of these
were included is that then the factor $1/2$ in Eq. (\ref{gammaxi})
should be removed. However, the symmetric structure of
Eq. (\ref{inteqxi}) means that
\begin{equation}
\Xi_k(q,p_1,p_2) = \Xi_k(q,p_2,p_1)
\label{eq:symmetry}
\end{equation}
which will be useful in the solution of these equations. How we solve
the set of equations (\ref{inteqgammas}),(\ref{inteqxi}), and
(\ref{gammaxi}) is described in detail in Appendix \ref{app:scat}.

As in the case of fermion-boson scattering, to calculate the
scattering amplitude each external bosonic leg has to be renormalized
by $\sqrt Z=\sqrt{8\pi/(m^2a)}$. The scattering amplitude is then
evaluated on shell
\begin{equation}
T^{\rm bb}(k) = Z^2 t^{\rm bb}_k(k,0).
\end{equation}
with $k\equiv |\vec k|$. Here units are restored to the
$t$-matrix. The scattering amplitude is related to the boson-boson
scattering length by
\begin{equation}
T^{\rm bb}(0) = \frac{2\pi}ma_b.
\end{equation}
By solving Eq. (\ref{inteqgammas}) using the method described above
and in Appendix \ref{app:scat} we find $a_b\approx0.60a$ in complete
agreement with Refs. \cite{Petrov2005,Brodsky2005}.

Using the above formalism it may also be checked that there are no
bound states of a pair of bosons. If such bound states exist, the
interaction between the bosons in this theory can become
effectively attractive, rendering the bosonic gas unstable. A bound
state of two bosons would correspond to a pole in the scattering
amplitude, that is, to a solution of the homogenous version of the
integral equation (\ref{inteqgamma}) for the $t$-matrix. Varying the
total energy, that is $k^2$ for $k^2\leq0$, we do not find any
solutions to the homogenous equation and thus do not find any bound
states.

It should be mentioned that the interaction between bosons does not
have to be renormalized, in contrast to the interaction between bound
states of bosons \cite{Efimov1971}. In the latter case, the
interaction between a boson and a dimer of bosons depends on the
ultraviolet cutoff and a possible solution to this difficulty is to
add a three body force counterterm to the theory and use an additional
input, such as an experimentally measured three body scattering length
\cite{Bedaque1998}. However, this difficulty is absent in the case of
scattering between a fermion and a dimer of fermions and also in the
problem of scattering between two dimers of fermions.

\section{Conclusion}

In this paper we discussed the low density expansion of the BCS-BEC
condensate in the BEC regime. We found the dispersion of its
Bogoliubov modes (the speed of sound), its chemical potential, and
the condensate depletion in the lowest order approximation in powers
of the gas parameter $a n^{\frac{1}{3}}$. Notice that the gas
parameter increases as the system is tuned towards the BCS-BEC
crossover regime, eventually reaching infinity at the so-called
unitary point lying in between the BCS and BEC condensate. So the
theory developed here works in the BEC regime only and breaks down
in the crossover area.

Throughout the paper we emphasized that the BCS-BEC gap equation
actually breaks down in the BEC regime. We would like to remark
further on the origin of this breakdown. The gap equation is derived
by minimizing the effective action of the condensate, given by
\rfs{eq:Seff1}. This is correct only if the fluctuations about this
minimum are small. If they are not small, the fluctuations must be
taken into account. In that case, minimization of the effective action
should be replaced by minimization of the effective potential, which
takes into account fluctuations, as discussed in any textbook on field
theory. In practice, it is possible to replace the evaluation of the
minimum of the effective potential by the Hugenholtz-Pines relation,
as is done here. The advantage of this second procedure is in the fact
that it automatically gives the excitation spectrum in addition to the
chemical potential. The two techniques are however equivalent. Most
importantly, the calculations performed in this paper demonstrate that
in including the fluctuations into the effective potential, one needs
to go well beyond the Gaussian fluctuations approximation. In fact,
terms up to infinite order (in terms of the number of loops) have to
be summed up, all of them being of the same order in the gas
parameter.  Fortunately, this is possible to do, and this is what has
been accomplished in this paper. Alternatively,
the gap equation corrected by fluctuations was calculated in
section~\ref{sec:BCS}.
We demonstrated how the naive gap equation breaks down due to
fluctuations, and how the fluctuations effectively replace the Born
approximation with the full boson-boson scattering amplitude. If we
wanted, we could follow the procedure outlined in
section~\ref{sec:BCS} as an alternative to the Hugenholtz-Pines
relation.

The same argument also demonstrates that the Gross-Pitaevskii equation
of the condensate should follow not from the effective action, but
from the effective potential. In other words, the true bosonic
scattering length \rfs{eq:shlyap} should be used in its quartic term,
as opposed to its Born approximation value $2 a_b$.  The
Gross-Pitaevskii equation will of course be valid only at length
scales much bigger than $a$.

All calculations in this paper are done in the lowest approximation
in density, which significantly simplifies the work needed to be
done. Finding higher order corrections to \rfs{eq:results} is one
possible direction of further work along the lines discussed in this
paper. Another possible direction is to study the condensate at
finite temperature. This should probably be done in the large $N$
approximation such as the one used in Ref.~\cite{Baym2000}.
Generalizing the techniques of our paper to finite temperature
should be a promising direction of further research.

On the other hand, we do not expect that these techniques will help
to shed light on the BCS-BEC crossover regime, especially at the
unitary point. The small parameter utilized here becomes infinity
there. The unitary regime can only be understood numerically
\cite{Bulgac2005}, barring an invention of an exact solution.

\begin{acknowledgments}
We thank J. Shepard, L. Radzihovsky, D. Sheehy, and A. Lamacraft for
useful discussions. This work was supported by the NSF grant
DMR-0449521.  J.~L. also wishes to thank the Danish Research Agency
for support.
\end{acknowledgments}

\appendix

\section{The Two channel model}
\label{AppendixTC}  The two channel model is defined by its
functional integral
\begin{equation} \label{eq:tc1}
Z=\int \cD \bar \psi_\uparrow \cD \psi_\uparrow \cD \bar
\psi_\downarrow \cD \psi_\downarrow \cD b \cD \bar b~e^{iS_{\rm
tc}},
\end{equation}
where the action $S_{\rm tc}$ is  given by $S_{\rm
tc}=S_0+S_{0b}+S_{ab}$. Here $S_0$ is the free fermion action given
in \rfs{eq:S}. $S_{0b}$ is the free boson action given by
\begin{equation} \label{eq:Sb}
S_{0b}= \int d^3x~ dt~ \bar b \left( i
\pp{t}+\frac{1}{4m}\frac{\d^2}{\d {\bf x}^2}+2\mu -\epsilon_0\right)
b,
\end{equation}
where $\epsilon_0$ is the detuning, the parameter controlled by the
magnetic field in the Feshbach resonance setup. $S_{ab}$ is the
interaction term
\begin{equation}
S_{ab}= -g \int d^3 x~dt ~ \left( b \, \bar \psi_{\uparrow} \bar
\psi_{\downarrow} + \bar b \, \psi_{\downarrow} \psi_{\uparrow}
\right).\end{equation} The scattering of fermions  in the two
channel model is characterized by the scattering length $a$ and
effective range $r_0$ given by \cite{Andreev2004}
\begin{equation}
a=-\frac{m g^2}{4 \pi \left(\epsilon_0-\frac{g^2 m \Lambda}{2 \pi^2}
\right)}, \ r_0=-\frac{8\pi}{m^2 g^2}.
\end{equation}
In order to ensure that we work in the wide resonance regime $|r_0|
\ll n^{-{\frac{1}{3}}}$, $|r_0| \ll |a|$, in this paper we need to
take the limit $g \rightarrow \infty$, while simultaneously
adjusting $\epsilon_0$ so that $a$ remains finite. Introducing the
notation $\Delta=g b$, $\bar \Delta=g \bar b$, we find that $g$
disappears from $S_{ab}$, while $S_{0b}$, in the large $g$ limit,
becomes
\begin{equation}
S_{0b} = \left(\frac{m}{4 \pi a} - \frac{m \Lambda}{2 \pi^2}
\right)\int d^3 x dt \, \bar \Delta \Delta.
\end{equation}
We recognize the combination of parameters in brackets as
$-1/\lambda$, thanks to \rfs{eq:aforlambda}. Thus the two channel
model reduces to the one channel model as given by \rfs{eq:S} and
\rfs{eq:HS}.

One important lesson which follows from this discussion is that in the
wide resonance regime, it is not justified to use $g$ as a small
parameter and construct perturbative expansion in its powers.  Indeed,
$g$ is not only not small, it should actually be taken to infinity.

\section{Fermionic Scattering and renormalized bosonic propagator}
\label{AppendixB}
\begin{figure}[hbt]
\includegraphics[height=.35 in]{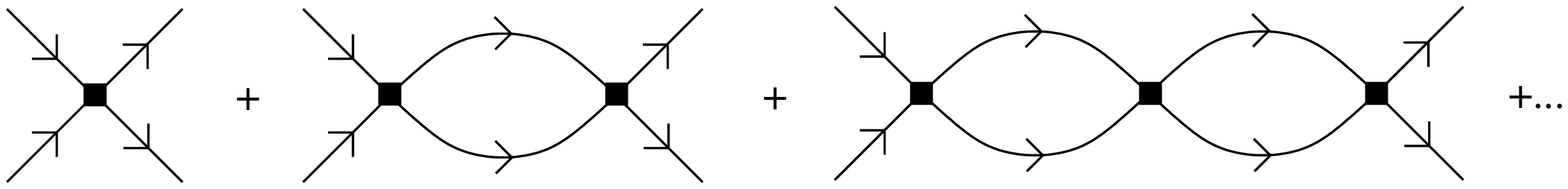}
\caption{The bubble diagrams of fermion scattering.}
\label{fig:ff}
\end{figure}
The scattering of fermions governed by the action Eq. (\ref{eq:S})
and interacting via the short range interaction Eq. (\ref{eq:Sint})
can be calculated by summing up the diagrams depicted in Fig.
\ref{fig:ff}. The calculation proceeds in vacuum, so the chemical
potential $\mu$ is set to zero everywhere. This sum gives the
fermion $T$-matrix, $T^{\rm ff}$, and the scattering length is
proportional to the $T$-matrix at zero momentum,
\begin{equation}
a = \frac m{4\pi}T^{\rm ff}(0).
\end{equation}
The bubble diagrams form a geometric series which can be summed to
give
\begin{equation}
T^{\rm ff}(0) = \frac{-\lambda}{1+\lambda\Pi(0)}
\end{equation}
where $\Pi$ is a bubble as shown in Fig. \ref{fig:ff} evaluated at
zero momentum. This bubble is given by
\begin{equation}
\Pi(0)  =  i\int\frac{d^4p}{(2\pi)^4}G_0(p)G_0(-p) \nn \\
=  -\frac{m\Lambda}{2\pi^2}.
\end{equation}
Here, $\Lambda$ is the momentum cut-off. This shows
Eq. (\ref{eq:aforlambda}), namely that
\begin{equation}
a = \left(-\frac{4\pi}{m\lambda}+\frac{2\Lambda}\pi\right)^{-1}.
\end{equation}

\begin{figure}[hbt]
\includegraphics[height=.9 in]{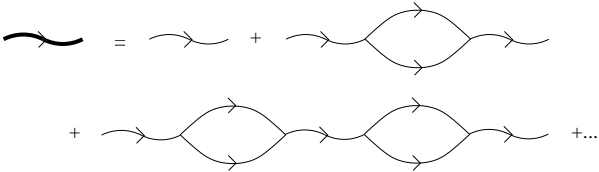}
\caption{The bosonic propagator renormalized by fermion loops. The
  thin wavy lines are unrenormalized boson propagators.}
\label{fig:bosonprop}
\end{figure}
The calculation of the bosonic propagator corrected by fermionic
loops proceeds similarly. Having in mind its applications in this
paper, we do this calculation at a finite chemical potential $\mu
\le 0$. Expanding the action $S_\Delta$ in Eq. (\ref{eq:Seff1})
results in a geometric series as shown in Fig. \ref{fig:bosonprop},
which can be summed to give
\begin{equation}
D_0(p) = \frac{-\lambda}{1+\lambda\Pi(p)}.
\label{eq:renormbos}
\end{equation}
The fermion loop can be calculated to give
\begin{eqnarray} \label{eq:polarization}
\Pi(p) & = & i\int\frac{d^4q}{(2\pi)^4}G_0(p+q)G_0(-q) \nn \\
&& \hspace{-1cm}=-\frac{m\Lambda}{2\pi^2}+\frac{m^{3/2}}{4\pi}\sqrt{-\omega+p^2/4m-2\mu-i0}.
\end{eqnarray}
Inserting this result in Eq. (\ref{eq:renormbos}), using Eq.
(\ref{eq:aforlambda}) to write the momentum cut-off in terms of $a$,
finally gives
\begin{equation} \label{eq:bosonicpropagator}
D_0(p) = \frac{4\pi}m\frac1{a^{-1}-\sqrt m\sqrt{-\omega+\frac{p^2}{4m}-2\mu-i0}}.
\end{equation}

Note that it is crucial that $\mu \le 0$ for the integral in
\rfs{eq:polarization} to be calculated the way it is. Otherwise, the
fermionic propagator is no longer retarded due to hole propagation.
Thus \rfs{eq:bosonicpropagator} applies only in vacuum or in the BEC
regime where $\mu \le 0$.

\section{The integral equation for boson-boson scattering}
\label{app:scat}
In this appendix we describe the solution of
Eqs. (\ref{inteqgammas}-\ref{gammaxi}), describing scattering of two
bound states, each consisting of two fermions. We demonstrate how to
integrate out the loop energies in constructing the sum of two boson
irreducible diagrams, $\Gamma$, and we write the corresponding
equations in a more convenient way for numerical studies. We also
describe the numerical methods used. For simplicity, in this appendix
all momenta are measured in units of $a^{-1}$, all energies in units
of $1/(ma^2)$, and $t^{\rm bb}$ and $\Gamma$ in units of $m^3a^3$.

\begin{figure}[hbt]
\includegraphics[height=1.8 in]{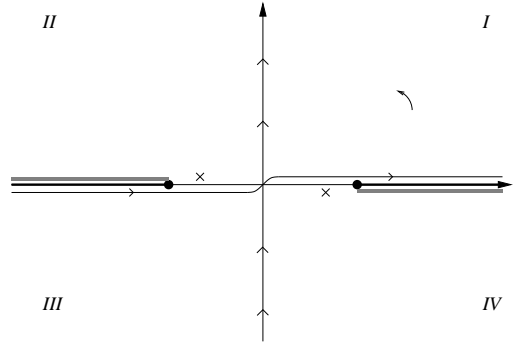}
\caption{The original and rotated contours of frequency integration used
  in Eq. (\ref{inteqgammas}). The crosses correspond to poles of the
  bosonic propagators which approach the imaginary axis as $k,q\to0$.
  The dots and black lines towards $\pm\infty$ are branch cuts of the
  bosonic propagators and the grey boxes are possible non-analytic
  structure of $\Gamma(q,q_0;p,p_0)$. These do not come closer to the
  imaginary axis than $\pm b$.}
\label{fig:contour}
\end{figure}
To construct $\Gamma_k(q,q_0;p,p_0)$ it is convenient first to rotate
the external energies $p_0$ and $q_0$ onto the imaginary axis which
can be done without crossing any poles. Both applications we have in
mind, calculation of the scattering length and search for bound
states, require the $t$-matrix $t^{\rm bb}_k(p,p_0)$ evaluated on
shell and thus we are only interested in the $t$-matrix evaluated at
$p_0=0$ which is still on the contour.  That it is possible to rotate
the external energies is not immediately obvious. Consider
Eq. (\ref{inteqgammas}) which relates the $t$-matrix to $\Gamma$. The
two bosonic propagators have branch cuts in $q_0$ starting at
$q_0=\pm(1+q^2/4-k^2/4)$ and going towards $\pm\infty$,
respectively. $1+q^2/4-k^2/4>0$ since to calculate the scattering
length we let $k=0$ and to search for bound states we let
$k^2\leq0$. The bosonic propagators also have poles at
$q_0=\pm(q^2/4-k^2/4-i0)$ where the infinitesimal is determined from
the requirement that Eq. (\ref{eq:bosonprop}) describes a retarded
propagator. It will be shown below that also $\Gamma(q,q_0;p,p_0)$ can
only be non-analytic in quadrant $II$ above the branch cut and $IV$
below the branch cut in $p_0$ and $q_0$. We conclude that not only is
it possible to rotate the external energies, but we also move the
contour of integration away from any singularities. This is
illustrated in Fig. \ref{fig:contour}. The only remaining
singularities close to the integration contour are from the poles of
the bosonic propagator as $q\to0$ and $k\to0$. This singularity is
integrable and we will explicitly treat this below by a change of
variables in Eq. (\ref{eq:change}).

In the integral equation (\ref{inteqxi}) it is possible
to integrate over the loop energy since $\Xi_k(q,p_1,p_2)$ is analytic
in the lower half planes of both fermion energies. To see this, note
that for any diagram contributing to $\Xi$ the two outgoing fermionic
propagators do not originate from the same bosonic propagator since
then $\Xi$ would be two boson reducible. Instead each of the fermions
originate from a boson which in turn decays into one other fermion
which through a series of propagations forward in time contribute to
the creation of the outgoing boson. Thus $\Xi$ contains only retarded
propagators of $-p_1$ and $-p_2$ and must be analytic in the lower
half planes of the corresponding energies.

It is not as simple to integrate out the loop energy in
Eq. (\ref{gammaxi}) since $\Xi_k(q,q',p-q')$ is not analytic in either
$q'_0$ half planes. However, if one considers the integral equation
satisfied by $\Xi_k(q,q',p-q')$, Eq. (\ref{inteqxi}) with $p_1=q'$ and
$p_2=p-q'$, the first product of propagators can be split into a term
analytic in the upper $q'_0$ half plane and a term analytic in the
lower $q'_0$ half plane by noting that
\begin{eqnarray}
G_0(K/2+q-q')G_0(K/2-q-p+q') \nn \\
= \frac{G_0(K/2+q-q')+G_0(K/2-q-p+q')}
{\frac E2-p_0-(\vec q-\vec q\,')^2/2-(\vec q+\vec p-\vec q\,')^2/2+i0}.
\end{eqnarray}
The second product of fermionic propagators is treated similarly.
The remaining part of the integral equation for $\Xi_k(q,q',p-q')$
consists of a term analytic in the upper $q'_0$ half plane and a term
analytic in the lower $q'_0$ half plane. Thus effectively we have
split $\Xi_k(q,q',p-q')$ into
\begin{equation}
\Xi_k(q,q',p-q') = \Xi^+_k(q,q',p-q') + \Xi^-_k(q,q',p-q')
\end{equation}
with $\Xi^+_k(q,q',p-q')$ ($\Xi^-_k(q,q',p-q')$) analytic in the upper
(lower) $q_0'$ half plane. Note that this splitting means that in
complete generality
\begin{equation}
\Xi_k(q,p_1,p_2) = \Xi^+_k(q,p_1,p_2) + \Xi^-_k(q,p_1,p_2).
\label{xisplit}
\end{equation}

$\Xi^+_k(q,q',p-q')$ and $\Xi^-_k(q,q',p-q')$ satisfy a set of coupled
integral equations. The equation satisfied by $\Xi^-_k(q,q',p-q')$ is
\begin{widetext}
\begin{eqnarray}
&& \Xi^-_k(q,q',p-q') =
\nn \\ &&
-\int\frac{d\Omega_{\vec q}}{4\pi}
\left[\frac{G_0(K/2+q-q')}
{\frac E2-p_0-(\vec q-\vec q\,')^2/2-(\vec q+\vec p-\vec q\,')^2/2+i0}
+\frac{G_0(K/2-q-q')}
{\frac E2-p_0-(\vec q+\vec q\,')^2/2-(\vec q-\vec p+\vec q\,')^2/2+i0}
\right] \nn \\ &&
-\left.\int\frac{d^3Q}{(2\pi)^3}G_0(K/2-Q-p)D_0(K-Q-q')\Xi_k(q,q',Q)\right|_{Q_0=\vec
  Q^2/2-E/4}.
\label{ximinusoff}
\end{eqnarray}
\end{widetext}
The right hand side of this equation is related to $\Xi^+$ through
Eq. (\ref{xisplit}). The crucial point is that the right hand side of
Eq. (\ref{ximinusoff}) equals $\Xi^+_k(q,p-q',q')$ which can be seen
by writing the corresponding equation for $\Xi^+_k(q,q',p-q')$ and
using the change of variables $p-q'\leftrightarrow q'$ along with
Eq. (\ref{eq:symmetry}). Thus
$\Xi^+_k(q,p-q',q')=\Xi^-_k(q,q',p-q')$. Consequently, using
Eq. (\ref{gammaxi}), we conclude that $\Xi^+_k(q,q',p-q')$ and
$\Xi^-_k(q,q',p-q')$ has identical contributions to $\Gamma$. We also
note that
\begin{equation}
\Xi_k(q,p_1,p_2) = \Xi^-_k(q,p_1,p_2) + \Xi^-_k(q,p_2,p_1),
\label{xisplitminus}
\end{equation}
which may be inserted in the right hand side of
Eq. (\ref{ximinusoff}). The result is an un-coupled integral equation
for $\Xi^-_k(q,q',p-q')$.

We now perform the $q'_0$-integration in the lower half plane in
Eq. (\ref{gammaxi}) using Eq. (\ref{xisplitminus}). This integration
sets $q'_0=\vec q\,'^2/2-E/4>0$. At this value of the total outgoing
fermion energy Eq. (\ref{ximinusoff}) reduces to
\begin{widetext}
\begin{eqnarray}
&& \Xi_k^-(q;\vec q\,',\vec q\,'^2/2-E/4;\vec p-\vec q\,',p_0-\vec
q\,'^2/2+E/4)
\nn \\ &=&
\int\frac{d\Omega_{\vec q}}{4\pi}
\left[
\frac{G_0(K/2+q-q')}
{p_0+(\vec q-\vec q\,')^2/2+(\vec q+\vec p-\vec q\,')^2/2-E/2-i0}+
\frac{G_0(K/2-q-q')}
{p_0+(\vec q+\vec q\,')^2/2+(-\vec q+\vec p-\vec q\,')^2/2-E/2-i0}
\right]
\nn \\
&& +\frac1{2\pi^2}\int d^3Q\,\frac
{\Xi^-_k(q;\vec q\,',\vec q\,'^2/2-E/4;\vec Q,\vec Q/2-E/4)
+\Xi^-_k(q;\vec Q,\vec Q/2-E/4;\vec q\,',\vec q\,'^2/2-E/4)}
{\left(p_0+\vec Q^2/2+(\vec p+\vec Q)^2/2-E/2-i0\right)\left(1-\sqrt{-E+3\vec
    Q^2/4+3\vec q\,'^2/4+\vec Q\cdot \vec q\,'/2}\right)}.
\label{ximinus}
\end{eqnarray}
\end{widetext}
The right hand side of this equation only contains $\Xi^-$ evaluated
at fermion energies related to the corresponding momentum by $Q_0=\vec
Q^2-E/4$ and $q'_0=\vec q\,'^2/2-E/4$, which we will call ``on
shell''. We will solve the integral equation by first letting
\begin{equation}
p_0\to \vec p\,^2/2+\vec q\,'^2-\vec p\cdot\vec q\,'-E/2,
\end{equation}
which means that the left hand side is also evaluated ``on shell''.
The resulting integral equation is an integral equation in $|\vec
q'|$, $|\vec p-\vec q\,'|$, and the angle between these, using $|\vec
k|$, $|\vec p|$, $|\vec q|$, and $q_0$ as input. Subsequently the ``on
shell'' $\Xi^-$ can be inserted in the right hand side of
Eq. (\ref{ximinus}) to evaluate $\Xi^-$ at any values of the outgoing
energy.

It is now possible to conclude that $\Gamma(q,q_0;p,p_0)$ is analytic
in quadrant $I$ and $III$ in $p_0$ and $q_0$ and that furthermore any
non-analyticity is far from the rotated contour of integration. The
remaining pole of a fermionic propagator in Eq. (\ref{gammaxi}) after
the $q_0'$ integration in the lower half plane is at $p_0=1-k^2/4+\vec
q\,'^2/2+(\vec p-\vec q\,')^2/2-i0$ where the real part is greater
than $1-k^2/4\geq1$. In Eq. (\ref{ximinus}) the poles in $p_0$ are in
quadrant $II$ and also has real part less than
$-1+k^2/4\leq-1$. Finally, also in Eq. (\ref{ximinus}), the pole in
$q_0$ is in quadrant $IV$ below the branch cut.

We now define the ``on shell'' function
\begin{eqnarray}
\tilde\Xi_k(q;p_1,p_2,\cos\theta) & \equiv &
\nn\\ &&
\hspace{-2.2cm}
\Xi^-_k(q;\vec p_1,\vec
p_1\,^2/2-E/4; \vec p_2,\vec p_2\,^2-E/4).
\label{eq:defineonshell}
\end{eqnarray}
Here, $p_{1,2} \equiv |\vec p_{1,2}|$ and $\theta$ is the angle
between these vectors. We then find the final set of equations to be
solved to construct $\Gamma$:
\begin{widetext}
\begin{eqnarray}
\tilde\Xi_k(q;p_1,p_2,\cos\theta) & = & -\int\frac{d\Omega_{\vec
    q}}{4\pi}
\left[
\frac1{q^2+p_1^2+p^2_2-\vec q\cdot \vec p_1+\vec q\cdot\vec p_2-E}
\frac1{-q_0+p_1^2/2+(\vec q-\vec p_1)^2/2-E/2} \right.
\nn \\
&&
\hspace{-1.8cm}
\left.+
\frac1{q^2+p_1^2+p^2_2+\vec q\cdot \vec p_1-\vec q\cdot\vec p_2-E}
\frac1{q_0+p_1^2/2+(\vec q+\vec p_1)^2/2-E/2}
\right] \nn \\
&&
\hspace{-1.8cm}
+\frac1\pi\int_{-1}^1d(\cos\phi)\int_0^\infty Q^2dQ\frac{1}
{1-\sqrt{-E+3Q^2/4+3p_1^2/4+Qp_1\cos\phi/2}} \nn \\
&&
\hspace{-1.8cm}
\times\frac{\tilde\Xi_k(q,p_1,Q,\cos\phi)+\tilde\Xi_k(q;Q,p_1,\cos\phi)}
{\sqrt{(Q^2+p_1^2+p_2^2+Qp_1\cos\phi+p_1p_2\cos\theta+Qp_2\cos\theta\cos\phi-E)^2
-Q^2p_2^2(1-\cos^2\theta)(1-\cos^2\phi)}}.
\label{onshellxi}
\end{eqnarray}
We now insert Eq. (\ref{onshellxi}) in the right hand side of
Eq. (\ref{ximinus}) using Eq. (\ref{eq:defineonshell}), and
subsequently Eq. (\ref{ximinus}) in Eq. (\ref{gammaxi}). Performing
the integral over the azimuthal angle in the $\vec Q$ integral we find
\begin{eqnarray}
\Gamma_k(q,q_0;p,p_0) & = & \Gamma_k^{(0)}(q,q_0;p,p_0)+
\frac1{4\pi^3}\int_{-1}^1 d(\cos\theta)\int_0^\infty
\frac{q'^2dq'}{p_0-q'^2-p^2/2+p q'
\cos\theta+E/2} \nn \\ && \times \int_{-1}^1d(\cos\phi) \int_0^\infty
\frac{Q^2dQ}{1-\sqrt{-E+3Q^2/4+3q'^2/4+Q q'\cos\phi/2}}
\nn \\ && \times
\frac{\tilde\Xi_k(q;q',Q,\cos\phi)+
\tilde\Xi_k(q;Q,q',\cos\phi)}{
\sqrt{(Q^2+p^2/2+Qp\cos\theta\cos\phi+p_0-E/2)^2-Q^2p^2
    (1-\cos^2\theta)(1-\cos^2\phi)}},
\label{eq:finalgamma}
\end{eqnarray}
where $\Gamma^{(0)}$ is the contribution from the first diagram on the
right hand side in Fig. \ref{fig:diagrams}, which can be found to be
\begin{eqnarray}
\Gamma_k^{(0)}(q,q_0;p,p_0) & = & \frac1{16\pi^2}\int_0^\infty
\frac{dq'}{pq}\frac1{-E/2+q'^2+p^2/4+q^2/4}
\ln\frac{-q_0^2+(-E/2+q'^2+p^2/4+q^2/4-q'p)^2}{-q_0^2+(-E/2+q'^2+p^2/4+q^2/4+q'p)^2}
\nn \\ &&
\times\ln\frac{-p_0^2+(-E/2+q'^2+p^2/4+q^2/4-q'q)^2}{-p_0^2+(-E/2+q'^2+p^2/4+q^2/4+q'q)^2}.
\label{gamma0}
\end{eqnarray}
\end{widetext}
With the energies rotated to the imaginary axis all equations used to
find the $t$-matrix, Eq. (\ref{inteqgammas}) at $dq_0\to i\,dq_0$, and
Eqs. (\ref{onshellxi}-\ref{gamma0}) are real. This is not immediately
obvious but may be easily checked numerically.

To treat the poles in Eq. (\ref{inteqgammas}) of the
bosonic propagators at $q_0=\pm(q^2/4-k^2/4-i0)$ as $k,q\to0$ we
perform the following change of variables on the external energies and
momenta to ``polar'' coordinates
\begin{eqnarray}
q^2/4 & = & R^2_1\cos\theta_1, \hspace{1cm} q_0 = R_1^2\sin\theta_1,
\nn \\
p^2/4 & = & R^2_2\cos\theta_2, \hspace{1cm} p_0 = R_2^2\sin\theta_2,
\label{eq:change}
\end{eqnarray}
where $R_i\in[0,\infty[$ and $\theta_i\in[0,\pi/2]$. Here we restrict
the integration over $q_0$ to the upper imaginary axis, using the
symmetry of the integrand in Eq. (\ref{inteqgammas}) as
$q_0\to-q_0$. This symmetry follows from the symmetry of the
problem. For convergence reasons it is advantageous to integrate over
finite intervals, and to this end we use the change of variables
\begin{equation}
R_i = \frac2{z_i+1}-1.
\end{equation}
We employ the same change of variables on the internal loop momenta
$Q$ and $q'$ in Eqs. (\ref{onshellxi}) and (\ref{eq:finalgamma}),
\begin{equation}
Q = \frac2{z_Q+1}-1, \,\,\, q' = \frac2{z'+1}-1.
\end{equation}
To solve the integral equations (\ref{inteqgammas}) and
(\ref{onshellxi}) we use the Nystrom method, writing the integral
equations as matrix equations and inverting these to find the
solution. Evaluation of the integrals in these integral equations is
performed using Gauss-Legendre quadrature \cite{numrecipes}.

\bibliography{bec}

\end{document}